\documentclass[superscriptaddress,showkeys,prl,aps,twocolumn]{revtex4-1}
\usepackage{dcolumn}
\usepackage{bm}
\usepackage[USenglish]{babel}
\usepackage[colorlinks,hyperindex, urlcolor=blue, linkcolor=blue,citecolor=black, linkbordercolor={.7 .8 .8}]{hyperref}
\usepackage{graphicx}
\usepackage{float}
\usepackage{amsfonts}
\usepackage{amsmath}
\usepackage{amssymb}
\usepackage{amsbsy}
\usepackage{multirow}
\usepackage{nicefrac}

\begin{document}

\title {A non-invasive sub-surface electrical probe to encapsulated layers in van der Waals heterostructures}

\author{Mrityunjay Pandey}
\email{mrityunjay@iisc.ac.in}
\address{Centre for Nano Science and Engineering, Indian Institute of Science, Bangalore 560012, India}
\author{Radhika Soni}
\address{Department of Instrumentation and Applied Physics, Indian Institute of Science, Bangalore 560012, India}
 \author{Avi Mathur}
\address{Department of Instrumentation and Applied Physics, Indian Institute of Science, Bangalore 560012, India}
\author{Srinivasan Raghavan}
\address{Centre for Nano Science and Engineering, Indian Institute of Science, Bangalore 560012, India}
\author{U. Chandni}
\email{chandniu@iisc.ac.in}
\address{Department of Instrumentation and Applied Physics, Indian Institute of Science, Bangalore 560012, India}\email{chandniu@iisc.ac.in}

\begin{abstract}
van der Waals heterostructures formed by stacking different atomically thin layered materials have emerged as the sought-after device platform for electronic and optoelectronic applications. Determining the spatial extent of all the encapsulated components in such vertical stacks is key to optimal fabrication methods and improved device performance. Here we employ electrostatic force microscopy as a fast and non-invasive microscopic probe that provides compelling images of two dimensional layers buried over 30 nm below the sample surface. We demonstrate the versatility of the technique by studying heterojunctions comprising graphene, hexagonal boron nitride and transition metal dichalcogenides. Work function of each constituent layer acts as a unique fingerprint during imaging, thereby providing important insights into the charge environment, disorder, structural imperfections and doping profile. The technique holds great potential for gaining a comprehensive understanding of the quality, flatness as well as local electrical properties of buried layers in a large class of nanoscale materials and vertical heterostructures.
\end{abstract}


\maketitle
\section{I. Introduction}

Two dimensional (2D) layered materials such as graphene, hexagonal boron nitride (hBN) and transition metal dichalcogenides (TMDCs) offer a new platform to create vertical heterostructures, where different atomically thin layers can be picked and assembled deterministically to suit various niche applications. These vertical stacks, popularly known as van der Waals (vdW) heterostructures~\cite{vdW2} exhibit new chemical and physical phenomena that are drastically different from the constituent layers~\cite{Jariwala,Chandni,Dean,Wang, Kallol}. The first tested device geometry involved single layer graphene encapsulated within atomically flat, dangling bond-free hBN layers with reduced charge traps, giving rise to improved transport properties and near-theoretical electron mobilities~\cite{Dean,Wang}. The device fabrication protocols have improved substantially over the years with heterostructures routinely containing five or more atomically thin layers of different materials, with precise twist angles~\cite{Kim_NL}. Such advances have led to the demonstration of exotic physical phenomena such as the Hofstader butterfly~\cite{Hunt} and magic-angle superconductivity~\cite{Cao}. The collective behaviour of the various heterostructure elements become relevant in these geometries, leading to superlattices, charge redistribution, modified structural properties and twist angle-dependent physical parameters~\cite{Cao, Hunt,Mishchenko}. Furthermore, encapsulating 2D layered materials between hBN layers also proves useful in studying many reactive layered materials such as phosphorene, magnetic materials such as chromium tri-iodide and various 2D topological insulators~\cite{Favron,Huang}. While encapsulation offers unparalleled performance advantage, it obscures the sample characterization and device fabrication methods as these samples are often exfoliated and assembled in gloveboxes.

Characterizing the spatial extent and charge distribution profile of the buried layers in a heterostructure is currently challenging. A combination of techniques such as optical microscopy, Raman mapping, scanning electron microscopy and atomic force microscopy (AFM) are commonly employed to identify the layered materials. Raman mapping is a particularly useful technique which in addition to identifying the constituent layers non-invasively, also provides crucial information on doping, edge profiles, lattice temperature, defects and strain~\cite{Anindya,Neumann}. However, the technique is hugely limited by the spatial resolution of the laser spot size which is usually of the order of 500 nm - 1 $\mu$m. Also, for large samples with lateral dimensions of the order of tens of microns, the image acquisition is a rather slow process, often running into few hours. Optical microscopy techniques such as dark field imaging and differential interference contrast imaging, and AFM topography scans provide sporadic benefits, limited by the thickness of the top-most layer (usually a thick, insulating hBN layer with thickness varying from 10 to 40 nm). Notably, these techniques do not provide any further sample specific information such as doping profile, which usually needs further invasive electrical transport measurements. A simple, fast and non-invasive method that can identify individual layers in a heterostructure, and at the same time provide useful information on the spatial distribution of charge carriers is desirable for enhanced device throughput.

 \begin{figure*}
\centering
\includegraphics[scale=1]{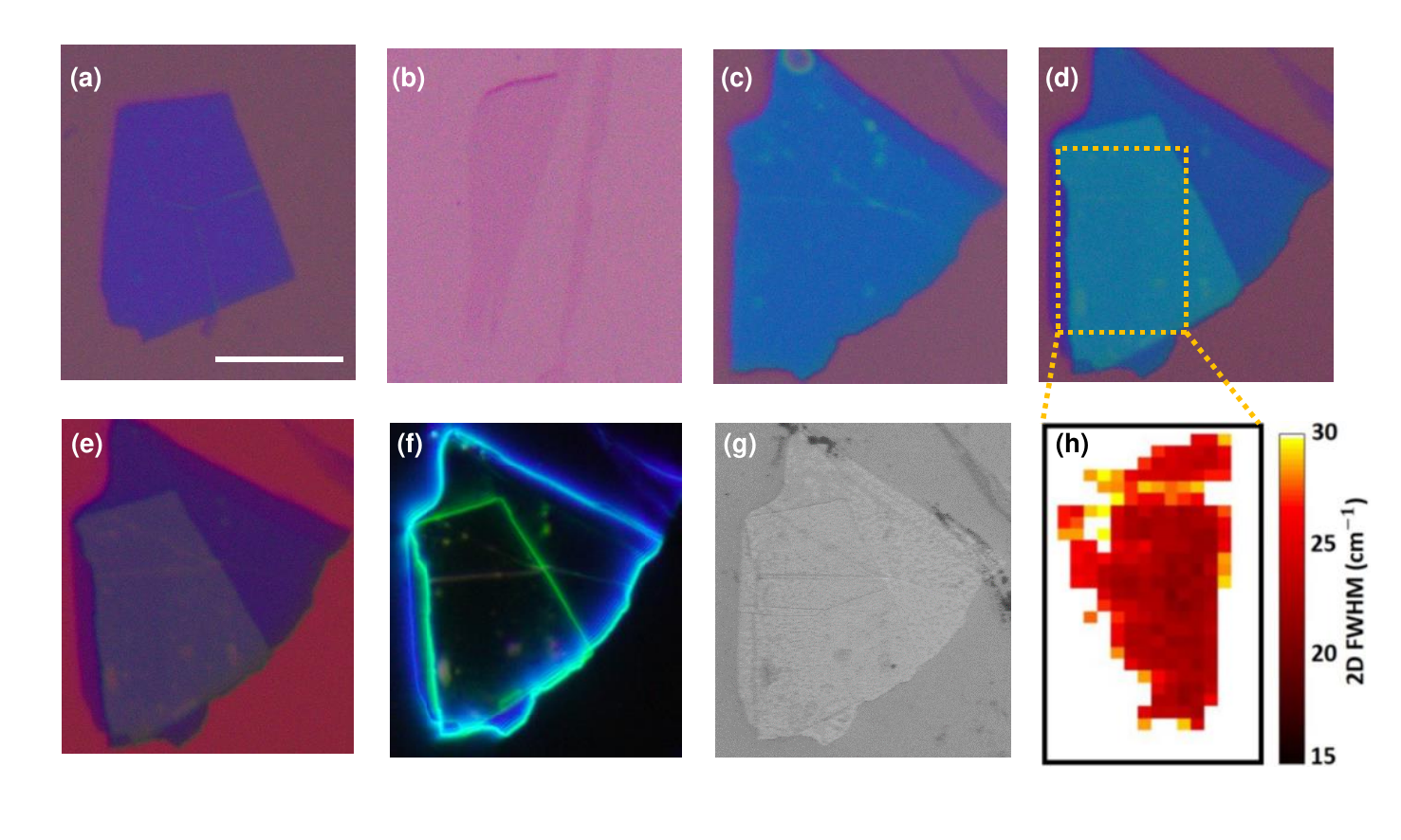}
\caption{Optical micrographs showing (a) top hBN, (b) single layer graphene and (c) bottom hBN, assembled into a vdW heterostructure S14, shown in (d). A differential interference contrast enhanced imaging mode is used in (e), while (f) shows a dark filed image of the same region. The scale bar is 10 $\mu$m. (g) Scanning electron microscopy image of the heterostructure taken at an accceleration voltage of 6 kV and working distance of 4 mm.  (h) Raman mapping of the 2D peak FWHM across the region shown by the yellow dashed lines, indicating the spatial extent of the encapsulated graphene flake.}
\end{figure*}


In this article, we present electrostatic force microscopy (EFM) as an effective tool to spatially map and study the local surface potential variations in encapsulated vdW heterostructures. In the past, surface potential profiles in isolated 2D layered materials have been investigated using EFM~\cite{Datta,Burnett,Panchal,Shi,Li_NL,Gomez} and Kelvin probe force microscopy (KPFM)~\cite{Yu,Yan,Druga,Ziegler,Willke,Robinson}. These studies were found to be particularly useful in studying doping~\cite{Shi}, substrate~\cite{Ziegler} and thickness dependences~\cite{Datta, Burnett} and contact resistances~\cite{Panchal,Yu,Yan,Druga,Willke} of graphene flakes supported on SiO$_2$/Si~\cite{Datta,Shi,Yu,Yan,Willke}, copper~\cite{Robinson} or SiC~\cite{Burnett, Panchal,Druga} substrates. Extension of this technique to vdW heterostructures involving multiple 2D layered materials has been limited~\cite{Li,Altvater}. Charge transfer mechanisms between the layers and screening of extrinsic charges from the substrate and external environment can signficantly alter the surface potentials when multiple 2D layers are stacked vertically. Hence, our current understanding of the surface potentials in isolated flakes do not work well in the context of complex, multi-layered heterostructures. Remarkably, EFM technique is capable of probing the individual layers in vdW heterostructures, which are otherwise inaccessible via conventional surface probes. In our measurements, the surface potential (or work function) acts as a fingerprint for each 2D layer, leading to a phase contrast with respect to other constituent layers. This enables us to obtain visually compelling images of various 2D layers, buried over 30 nm below the sample surface. 

We fabricate a variety of encapsulated heterostructures involving graphene, hBN and TMDCs, and demonstrate the ease of visually distinguishing the encapsulated 2D layers using EFM. We observe that the EFM phase as a function of tip-sample bias (or spectroscopy) provides a quantitative estimate of the work functions of the encapsulated layers. Further, we demonstrate the effect of substrate doping and ambient atmosphere on the work functions of graphene layers by fabricating and characterizing different heterostructure architectures and comparing them with fully encapsulated regions. Finally, we show that the technique has a broad appeal and works equally well for complex heterostructures involving multiple 2D layers. 

\section{II. Results and Discussions}

\begin{figure*}
\includegraphics[scale=0.45]{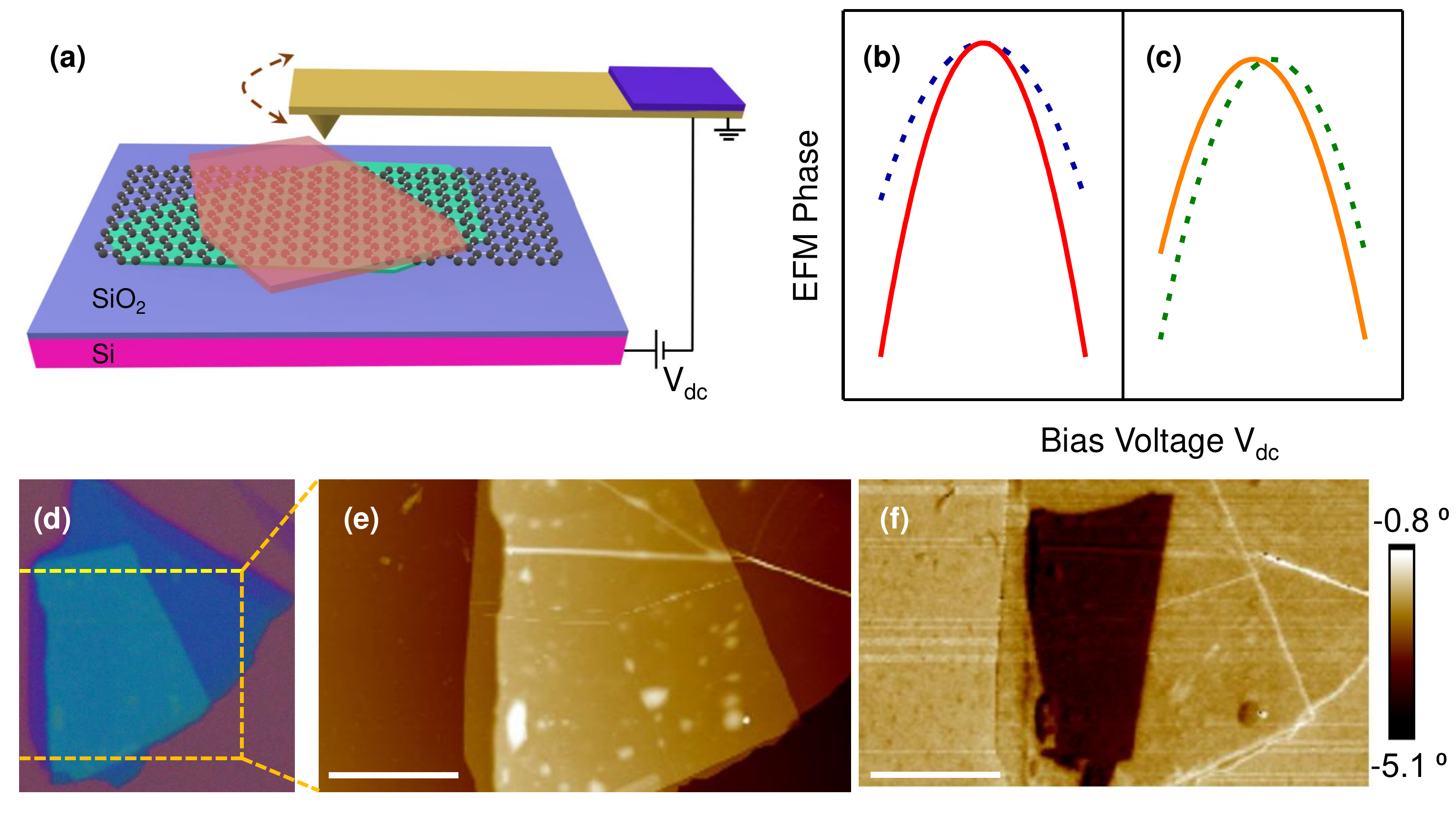}
\caption{(a) Schematic of the EFM setup, with a bias voltage $V_{dc}$ applied to the sample via the doped Si wafer and the tip acting as the ground terminal. Schematic showing two possible origins for the EFM phase contrast namely, (b) difference in the capacitive coupling between the heterostructure element and the AFM tip, leading to varied curvatures in EFM phase spectroscopy and (c) surface potential difference leading to offsets in the EFM phase parabola. (e) AFM topography for sample S14 shown in Fig.1, for the region marked in (d), showing wrinkles, bubbles and other topographic features. (f) EFM phase image for the same region at a sample bias of 1 V, clearly showing the encapsulated graphene layer. The scale bars are 6 $\mu$m.}
\end{figure*}

\subsection{A. Sample Fabrication and Characterization} 
Encapsulated vdW heterostructures comprising graphene, hBN and TMDCs were fabricated on SiO$_2$/Si substrates using the dry transfer method~\cite{Wang}. Individual graphene, hBN, MoS$_2$ and WSe$_2$ flakes were mechanically exfoliated on oxygen plasma treated Si substrates with a 285 nm SiO$_2$ top layer. The desired flakes were identified using an optical microscope. A Polypropylene carbonate (PPC)/Polydimethylsiloxane (PDMS) hemispherical stamp on a glass slide was used to pick up the top hBN at 45$^0$C, followed by graphene or MoS$_2$ or WSe$_2$. The stack was transferred on the exfoliated bottom hBN on SiO$_2$/Si wafer at 100$^0$C. To demonstrate the capability of the technique we first employed simple hBN/Graphene/hBN heterostructures. Fig. 1(a)-(c) show the constituent layers in such a stack (S14) where a single layer graphene (SLG) is encapsulated between hBN layers of thicknesses 13 nm (top) and 28 nm (bottom). Evidently, the optical image of the vertical stack shown in Fig. 1(d) does not reveal the encapsulated graphene flake. Even with a differential interference contrast optical microscopy (Fig. 1(e)), dark field imaging (Fig. 1(f)) and scanning electron microscopy (Fig. 1(g)), the encapsulated flake is not discernible. Further characterization of the encapsulated flake using Raman mapping (Fig. 1(h)) provided a reasonable estimate of the spatial extent. Raman mapping was done using a confocal microscope system ( LabRAM HR) with a 100x objective and a numerical aperture of 0.90. A 532 nm laser with a spot diameter of $\sim$ 1 $\mu$m was used to excite samples placed on a piezocrystal-controlled scanning stage. The Raman spectra were collected using a low laser power of 1.1 mW to avoid any laser induced heating effects. For the graphene samples, the 2D peak position and full width at half maximum (FWHM) were estimated using single-peak Lorentzian fits. Total acquisition time for the image shown in Fig. 1(h) was $\sim$ 4 hours.

\subsection{B. EFM as a non-invasive imaging tool} 
In Fig. 2, we present EFM measurements on stack S14. EFM measurements were conducted under ambient conditions using a Bruker Dimension Icon AFM setup employing Pt/Ir coated silicon probes (SCM-PIT V2) with a nominal tip radius of $\sim 25$ nm. EFM is a two-pass technique, where both topography and EFM phase are recorded sequentially. In the first pass, the AFM tip traces the topographic line profile in tapping mode. During the second pass, the AFM tip is lifted by a specified lift height $z$ and a $dc$ voltage $V_{dc}$ is applied to the sample via the doped Si wafer, with the AFM tip grounded (Fig. 2(a)). The tip is mechanically oscillated and retraces the topographic profile, providing a measure of the tip-sample electrostatic forces, given by $F=-\frac{1}{2}\frac{dC}{dz}(V_{dc}-V_s)^2$, where $V_s$ is the local electrostatic potential over the sample surface and $C$ is the tip-sample capacitance. $V_s=\Phi_s-\Phi_t$, also called the contact potential difference, can be used to estimate the work function (or electron affinity) of the sample ($\Phi_s$), provided the work function of the tip ($\Phi_t$) is known. The electrostatic force leads to a change in resonant frequency of the cantilever, which is measured as a phase shift, given by~\cite{Staii,Lei}:
\begin{equation}
\tan(\Delta\phi)=\frac{Q}{k}\frac{dF}{dz}=\frac{-Q}{2k}\frac{d^2C}{dz^2}(V_{dc}-V_s)^2
\end{equation}
where $Q$ and $k$ are the quality factor and force constant of the AFM probe respectively, which can be measured by the thermal tune calibration procedure. This gives a characteristic parabolic dependence of the EFM phase on the tip voltage.
Since the EFM technique operates on the force gradient, it has a much better sensitivity compared to amplitude modulated KPFM technique~\cite{Panchal}. In addition, the effect of the parasitic capacitances from the cantilever part of the AFM probe is minimised as the force gradient decays faster with distance. As can be seen from equation (1), a phase contrast can appear between constituent layers in a heterostructure due to two factors: (a) capacitive coupling, which leads to a change in the curvature of the EFM phase parabola and (b) surface potential $V_s$, which results in an offset in the EFM phase parabola. These two scenarios are depicted in Fig. 2(b) and Fig. 2(c) respectively. In a heterostructure geometry, we expect to have a combination of these two effects, leading to a phase contrast between various constituent layers. For the device S14, the AFM topography shown in Fig. 2(e) shows wrinkles, trapped bubbles and other topographical defects. However, the encapsulated SLG is not discernible.The EFM phase image for the same region, acquired at a bias of 1 V and a lift height of 30 nm is shown in Fig. 2(f). Remarkably, the encapsulated region and all the features therein, are visible with enhanced clarity in the EFM phase image. We attribute this contrast to the difference in the surface potential and the capacitive coupling in the constituent layers. Here, the top hBN layer, being an insulator with a large band gap of $\sim$ 5.9 eV, acts as an additional dielectric layer with a dielectric constant of $\epsilon_r\sim 4$ ~\cite{Li_NL}. The encapsulated semi-metallic graphene is hence visible with a good  contrast in the EFM phase image. These results establish EFM as a non-invasive, sub-surface imaging technique to detect encapsulated 2D layers in vdW heterostructures.

\begin{figure*}
\centering
\includegraphics[scale=0.7]{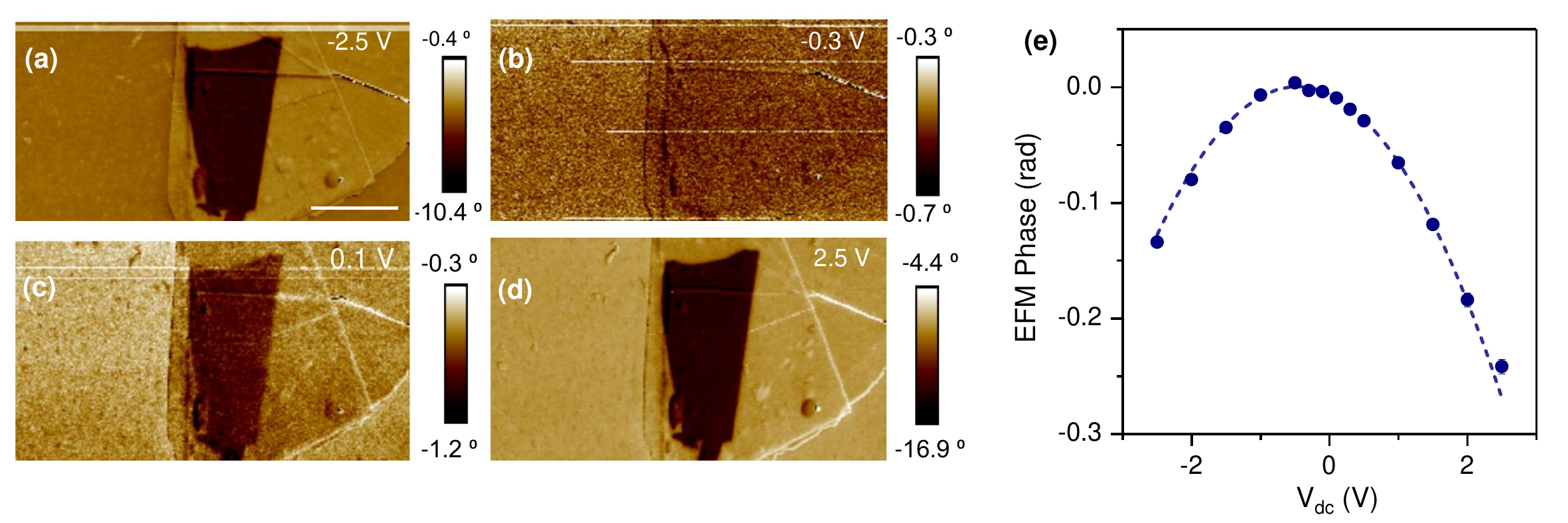}
\caption{EFM phase images for stack S14 shown in Fig. 1, at four different sample bias voltages: (a) -2.5 V,  (b) -0.3 V,  (c) 0.1 V and  (d) 2.5  V. The scale bar is $6~\mu$m.  (e) EFM phase as a function of sample bias for the encapsulated graphene flake. Fit to equation (2) is shown as a dashed line.}
\end{figure*}

In order to quantitatively analyse the surface potential of the encapsulated graphene layer, EFM phase images were taken at various sample bias voltages. In Fig. 3(a)-(d), we show the phase images at bias voltages $V_{dc}=-2.5$ V, -0.3 V, 0.1 V and 2.5 V respectively. We note that the phase shift of the graphene region was most often negative with respect to the background SiO$_2$ and hBN regions. While graphene is clearly visible at $\pm$ 2.5 V and 0.1 V, the contrast is poor for -0.3 V, where it merges with the SiO$_2$/hBN background phase. EFM phase vs. bias voltage on the graphene region (Fig. 3(e)) shows the characteristic parabolic dependence of equation (1). Fluctuations in phase within the graphene region were found to be minimal, evident from the negligible error bars. The maximum value in the parabola occurs when $V_{dc}=V_s$, which can be obtained by fitting the data with:

\begin{equation}
\Delta\phi=-\tan^{-1}[C_0(V_{dc}-V_s)^2]+C_1
\end{equation}
where the coefficient $C_0=(Q/2k)(d^2C/dz^2)$ and $C_1$ is the phase offset in the experiments. This offset has been corrected for clarity and ease of comparison in the EFM phase plots shown here.

\begin{figure*}
\centering
\includegraphics[scale=0.45]{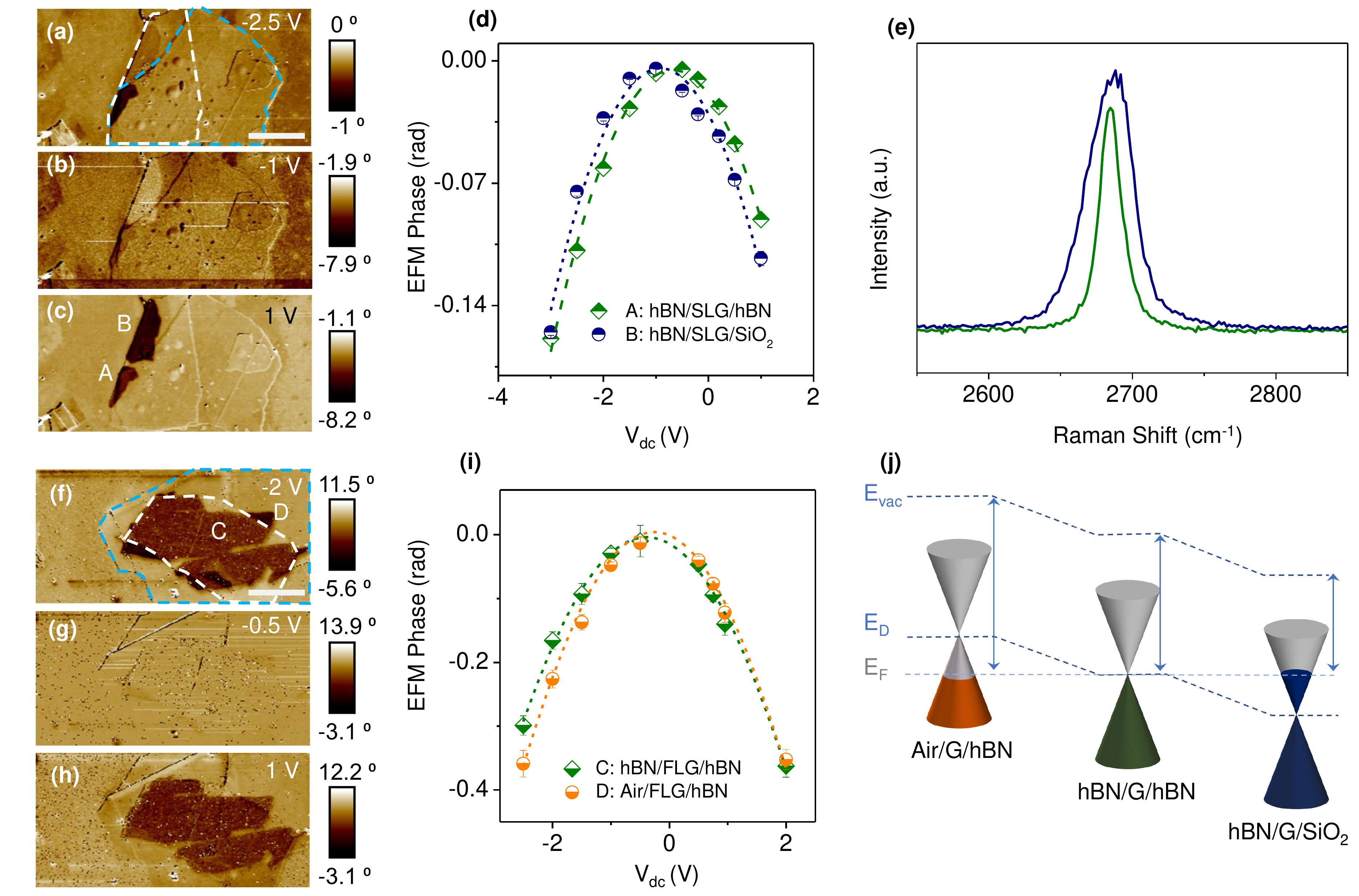}
\caption{(a)-(c) EFM phase images of stack S15, at bias voltages -2.5 V, -1 V and +1 V. In (a), white and blue dashed lines indicate the top and bottom hBN layers respectively. The hBN/SLG/hBN and hBN/SLG/SiO$_2$ regions indicated as A and B in (c), show distinct phase contrast. Scale bar is 8 $\mu$m. (d) EFM phase plots for regions A and B shown in (c). (e) Raman spectroscopy data showing the 2D peak intensity for the two regions A (green) and B (blue). (f)-(h) EFM phase images of a few layer graphene stack S8, at bias voltages -2 V, -0.5 V and 1 V. In (f), the white and blue dashed lines indicate the top and bottom hBN layers respectively. The hBN/FLG/hBN and Air/FLG/hBN regions indicated as C and D in (f), show distinct phase contrast, as seen in the EFM phase plot in (i). Illustration of the energy bands for three different heterostructure regions are shown in (j). $E_{\text{vac}}$, $E_F$ and $E_D$ denote the vacuum, Fermi and Dirac levels. Vertical arrows indicate work functions.}
\end{figure*}

The maximum point extracted from the parabolic fit gives a good estimate of the surface potential, which is obtained as $V_s = -0.46\pm0.03$ V for the encapsulated SLG. While $Q$ and $k$ can be obtained experimentally, the capacitive coupling is rather complex with contributions from the tip apex, cone and the cantilever~\cite{Lei}. For the encapsulated flake, the capacitive coupling is more involved, as the metal tip-semimetal capacitance is no longer passive and cannot be analytically determined from the tip-sample geometry alone. Due to these constraints on the quantitative estimate of the capacitive coupling of the tip and heterostructure elements, we restrict our analysis to the surface potential estimation alone. In order to quantify the work function of the constituent layers, it is important to define the work function of the AFM probe tip. This calibration was done using a gold sample ($\Phi_{\text{Au}}=5.1 $ eV), giving $\Phi_t=5.3$ eV, which is a good estimate for the Pt/Ir AFM probe. For sample S14, we obtain the work function to be $\sim$ 4.84 eV. \\

In order to demonstrate the versatility of the technique and its applicability to a larger class of 2D materials, we have also performed EFM measurements on encapsulated TMDCs, such as MoS$_2$ and WSe$_2$, the details of which can be found in the supplemental material section. Table 1 summarises the results over various encapsulated samples.

\begin{table}[t]
\centering
\begin{tabular}{|l|l|r|l|}
\hline
Type & $\text{Sample}$ & $V_s(V) $ & $\Phi_s (eV)$ \\
\hline
\multirow{2}{*}{hBN/SLG/hBN} &S14  & -0.46 & 4.84\\
&S15 & -0.66 & 4.64 \\

\hline
\multirow{2}{*}{hBN/FLG/hBN} &S8 & -0.39 & 4.91\\
&S11 & -0.45 & 4.85\\
\hline
hBN/FL-MoS$_2$/hBN & S21 &  -0.80 & 4.5 \\
\hline
hBN/FL-WSe$_2$/hBN & S20 &  -0.26 & 5.04\\
\hline
\end{tabular}
\caption{ Surface potential ($V_s$) and work function ($\Phi_s$) for different encapsulated 2D layers measured in this work. $V_s$ was estimated using fit to equation (2) and $\Phi_s$ was estimated using a calibrated AFM tip.}
\end{table}

\subsection{C. Probing doping in encapsulated layers using EFM}

In order to further demonstrate the capability of the technique in probing electrical properties of the buried layers, we utilize diverse heterostructure geometries. In Fig. 4(a)-(c), we show the EFM phase images for two SLG regions in stack S15, with slightly different vertical architectures, as shown by regions A and B in Fig. 4(c). While region A comprises a fully encapsulated graphene flake, in region B, the graphene flake is on SiO$_2$/Si wafer with a top hBN layer. The top and bottom hBN thicknesses measured by AFM topography were 18 nm and 33 nm respectively, and EFM phase mapping was performed at a lift height of 30 nm. Interestingly, we observe clear phase contrast in the two regions, the relative sign of which changes as the sample bias is varied. We also note that within the two regions, the fluctuation in phase is minimal. The regions are thus nearly equipotentials, with the surface potential varying only on changing the charge environment via the heterostructure geometry. 
 \begin{figure*}
\centering
\includegraphics[scale=0.45]{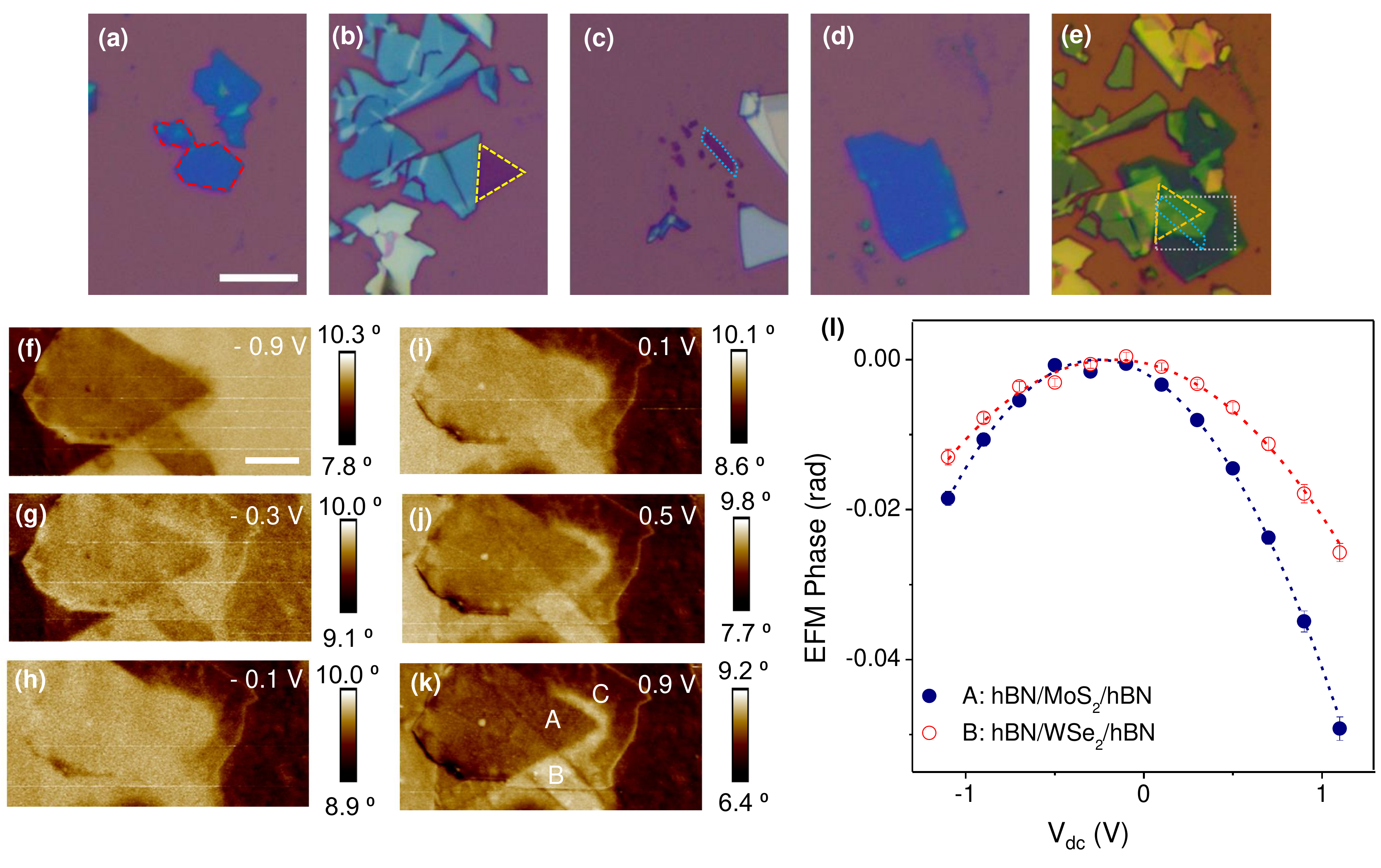}
\caption{Optical micrographs of the constituent layers in stack S18 showing (a) top hBN indicated by red dashed lines, (b) MoS$_2$ flake indicated by yellow dashed lines, (c) WSe$_2$ flake indicated by blue dashed lines (d) bottom hBN and (e) assembled heterostructure with the architecture hBN/MoS$_2$/WSe$_2$/hBN. Scale bar is 10 $\mu$m. Gray dashed box in (e) indicates the region over which EFM phase images (f)-(k) were acquired at various bias voltages. Phase image at - 0.1 V (h) shows least phase contrast between WSe$_2$ and MoS$_2$, while image at 0.9 V (k) shows clear phase contrast between encapsulated MoS$_2$ and WSe$_2$. At this voltage, top hBN (region C), MoS$_2$ (A) and WSe$_2$ (B) can easily be differentiated. Scale bar for EFM images is 2 $\mu$m. (l) EFM phase plots for two different regions A and B as marked in (k), comprising encapsulated MoS$_2$ and encapsulated WSe$_2$.}
\end{figure*}

To further understand the phase contrast, we have plotted EFM phase vs. voltage plots (Fig. 4(d)) for the two regions. The parabolic curves are found to be offset with respect to each other and cross over near $V_{dc}=-1$ V. This explains the changes in the phase contrast seen in Fig. 4(a)-(c). On fitting the two parabolas with equation (2), we obtain the surface potentials for the two configurations as   $V_s=-0.66\pm0.01$ V and $-1.06\pm0.05$ V for regions A and B respectively. We attribute this difference to doping from charge traps in the substrate. The encapsulated graphene regions (A) gave a work function $\approx 4.64$ eV, while region B gave lower ($\approx 4.24$ eV) value of work function, indicating electron doping. In region B, SiO$_2$ beneath the graphene has positive trap charges which modulate the chemical potential of graphene resulting in $n$-doping, similar to previous studies~\cite{Gomez,Romero}. In region A, the bottom hBN layer screens the effect of dangling bonds and trap charges of SiO$_2$, while the effect of the ambient environment on the chemical potential of graphene is reduced by the top hBN. The substrate induced doping is also verified by Raman spectroscopy measurements, where we obtain a broader 2D Raman peak in region B (Fig. 4(e)). Encapsulated region A has the 2D peak at $\sim$ 2684 cm$^{-1}$ with a FWHM of 15.7 cm$^{-1}$, while in region B, the peak is observed at $\sim$ 2685 cm $^{-1}$ with FWHM $\sim$ 30 cm$^{-1}$. Dielectric screening results in suppression of the electron-electron scattering in hBN supported graphene, giving a sharper peak in region A. In SiO$_2$  supported SLG, because of poor screening the scattering process is enhanced with lower life time of photoexcited electron-hole pairs, resulting in a broader 2D peak consistent with previous reports~\cite{Anindya}. 2D peak depends on the phonon life time as well as the number of scattering paths to satisfy the double resonance condition and is thus sensitive to inhomogeneity at small doping levels ($\sim 10^{12}$ cm$^{-2}$). The other factor that contributes to the electrostatic phase is the capacitive coupling between tip and sample, which should roughly be the same for both the regions. We note that the fitting indeed gives a comparable value for the coefficient $C_0$ to be 0.031 V$^{-2}$ and 0.026 V$^{-2}$ respectively, for the two regions A and B.

In Fig. 4(f)-(h), we show the EFM phase images for a few layer graphene (FLG) flake on hBN (stack S8), which is fully encapsulated in region C and exposed to air in region D. Top and bottom hBN flakes were 25 nm and 27 nm thick respectively, and the lift height for EFM measurements was 20 nm. Fig. 4(i) shows the EFM phase plots for the two regions. Again, we observe clear phase contrast in the two regions, with $V_s=-0.39\pm0.02$ V and $-0.23\pm0.02$ V for regions C and D respectively. The encapsulated graphene regions (C) gave a work function $\approx 4.9$ eV, while region D gave higher ($\approx 5.1$ eV) value of work function, indicating hole doping consistent with previous reports of surface potential mapping on few layer graphene films exposed to air~\cite{Datta}. The capacitive coupling is expected to be slightly higher in this measurement due to the reduced lift height, as validated by the coefficient $C_0=$ 0.067 and 0.072 V$^{-2}$ for regions C and D respectively. We note that variations in relative humidity, substrate charge environment and ambient atmosphere can modify the doping and lead to minor discrepancies across samples. However the overall trend remains as described above, in all samples measured (See supplemental material). Our observations are summarised in the energy band diagram shown in Fig. 4(j), assuming minimal doping in the fully encapsulated graphene region. 

\subsection{D. TMDCs-based heterostructures}
 We further demonstrate that the EFM phase elucidates the individual elements in complex heterostructures involving graphene, hBN and TMDCs. Such heterostructures are routinely used in many 2D-materials based optoelectronic devices~\cite{Kallol}, and the EFM technique can prove to be extremely useful in such studies as well. Fig. 5(a)-(d) show the optical micrographs of the constituents of stack S18 (Fig. 5(e)) consisting of hBN/MoS$_2$/WSe$_2$/hBN. The EFM phase images taken at bias voltages in the range $\pm 0.9$ V are shown in Fig. 5(f)-(k). We observe clear phase contrast across the device, as the bias voltage is varied. Two distinct regions consisting of encapsulated MoS$_2$ (region A) and encapsulated WSe$_2$ (region B) indicated in Fig. 5(k) are further explored in the phase plot shown in Fig. 5(l). As expected, the EFM phase parabolas are offset due to the difference in the surface potentials in the two layers. We obtain a value of $V_s=-0.26\pm 0.01$ V ($\Phi_s\approx 5.04 $ eV) and $-0.16\pm 0.01$ V ($\Phi_s\approx $ 5.14 eV) for regions A and B respectively. The work functions of the MoS$_2$ and WSe$_2$ layers in regions A and B are found to be slightly different from the encapsulated flakes reported in Table 1. We speculate this to be due to charge redistribution in the heterostructure. Earlier studies have reported similar charge redistribution and enhanced screening effect in single layer graphene films connected to few layer graphene films~\cite{Shi}. Studies have also found huge variations in the work function of TMDCs, strongly influenced by  thickness, substrates, charge environment and adsorbed layers~\cite{Yang,Feng,Jong}. We believe similar out-of-plane charge transfer mechanisms are possible in samples involving multiple 2D layers with overlapping regions, which will be explored in future studies. Nevertheless, the technique provides a reliable method to map elements in a complex heterostructure.

\section{III. Conclusions}

In summary, our results unambiguously demonstrate EFM phase measurements to be a useful tool not only in determining the spatial extent of encapsulated 2D flakes, but also in better understanding the charge distributions and surface potentials in various vertical heterostructures. We show that the 2D layers are affected by the ambient atmosphere and the charge traps in the substrate, which are easily identified non-invasively using EFM measurements. Furthermore, the technique provides very clear phase contrast between different 2D materials even in complex heterostructure architectures, which in conjunction with topography scans provide an easy fabrication route to high quality devices. We believe that these results will be valuable for diverse studies involving 2D materials and also greatly advance the creation of reliable and high throughput device architectures for electronic and optoelectronic applications.\\

\section{Supplemental Material}
Fig.6 shows additional data obtained for encapsulated WSe$_2$ and MoS$_2$ layers. Table II summarizes the results obtained for various graphene-based vdW heterostructures reported in this work. 
 \begin{figure*}
\centering
\includegraphics[scale=0.4]{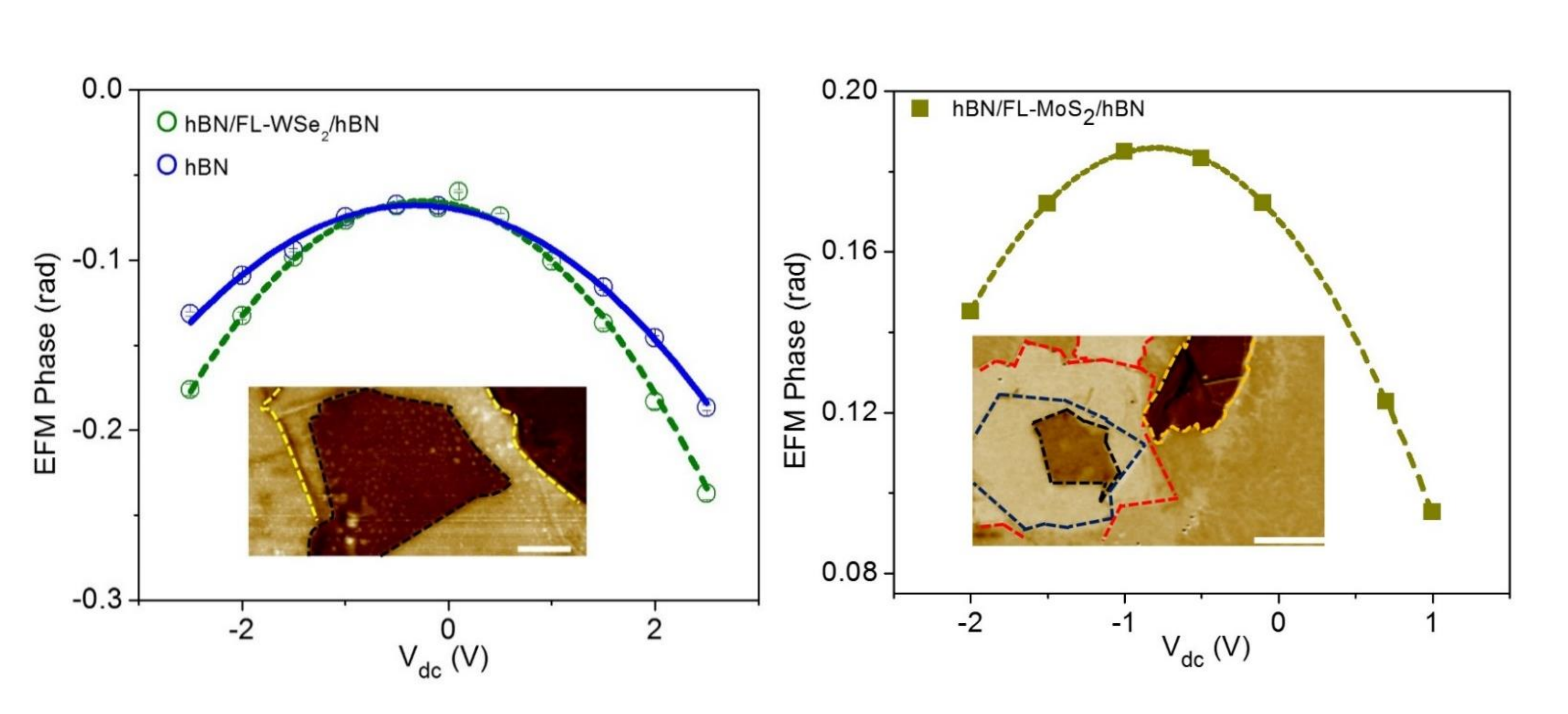}
\caption{(a) EFM Phase plot for an encapsulated FL-WSe$_2$ stack S20, showing clear distinction between the constituent hBN (blue) and WSe$_2$ (green) layers. Inset shows the EFM phase image at -2 V. Yellow and black dashed lines indicating boundaries of the top hBN and encapsulated WSe$_2$ respectively. The dashed lines indicate fit to equation 2 in the main text, giving the surface potential $V_s=-0.26$ V, and the coefficient $C_0$ = 0.022 V$^{-2}$ for the encapsulated WSe$_2$. Scale bar is 1 $\mu$m. (b) EFM Phase plot for an encapsulated FL-MoS$_2$ stack S21. Inset shows the EFM phase image at -2 V. Red, blue and black dashed lines indicate boundaries of the bottom hBN, top hBN and encapsulated FL-MoS$_2$ respectively. The darker region on the top right-hand side (yellow dashed lines) is a bulk MoS$_2$ flake. The dashed lines indicate fit to equation 2 in the main text, giving the surface potential V$_s=-0.8$ V, and the coefficient $C_o = 0.028 $V$^{-2}$. Scale bar is 4 $\mu$m.}
\end{figure*}

\begin{table*}
\centering
\begin{tabular}{|c|c|c|c|c|c|c|}
\hline
Type & Sample & Bottom hBN (nm) & Top hBN (nm ) & Lift Height (nm) & $C_o$ (V$^{-2}$) & $V_s$ (V) \\
\hline
hBN/SLG/hBN & S14 & 28 & 13 & 30 & 0.031 & -0.46 \\
\hline
hBN/SLG/SiO$_2$ & S15 & 33 & 18 & 30 & 0.026 & -1.06\\
\hline
hBN/SLG/hBN & S15 & 33 & 18 & 30 & 0.031 & -0.66 \\
\hline
hBN/FLG/hBN & S8 & 27 & 25 & 20 & 0.067 & -0.39 \\
\hline
Air/FLG/hBN & S8 & 27 & 25 & 20 & 0.072 & -0.23 \\
\hline
hBN/FLG/hBN & S11 & 27 & 27 & 30 & 0.040 & -0.45 \\
\hline
hBN/FLG/hBN & S11 & 27 & 27 & 30 & 0.050 & -0.36 \\
\hline
\hline\hline
hBN/FL-WSe$_2$/hBN & S20 & 24 & 18 & 30 & 0.022 & -0.26 \\
\hline
hBN/FL-MoS$_2$/hBN & S21 & 34 & 12 & 30 & 0.028 & -0.80\\
\hline
\end{tabular}
\caption{ Table summarizing the results obtained for various graphene and TMDC-based vdW heterostructures, indicating top and bottom hBN thicknesses, lift height and fit parameters $C_o$  and $V_s$ obtained from equation (2). Within a stack the variations in capacitive coupling is expected to be minimal. The major factor responsible for phase variance is the change in the surface potential of the constituent layers.}
\end{table*}

\section{Acknowledgements}
We thank T. Ahmed, S. Bhattacharyya, A. Ghosh, E. A. Henriksen and A. N. Pal for useful discussions. We gratefully acknowledge the usage of the Micro and Nano Characterization Facility (MNCF) at CeNSE, IISc and oxygen plasma system from OMI Laboratory, IISc. U.C. acknowledges support from the Infosys Foundation, IISc start-up grant and ISRO-STC cell.

%

\end{document}